# The causal foundations of applied probability and statistics

Sander Greenland
Department of Epidemiology and Department of Statistics,
University of California, Los Angeles
lesdomes@ucla.edu

**Abstract**. Statistical science (as opposed to mathematical statistics) involves far more than probability theory, for it requires realistic causal models of data generators – even for purely descriptive goals. Statistical decision theory requires more causality: Rational decisions are actions taken to minimize costs while maximizing benefits, and thus require explication of causes of loss and gain. Competent statistical practice thus integrates logic, context, and probability into scientific inference and decision using narratives filled with causality. This reality was seen and accounted for intuitively by the founders of modern statistics, but was not well recognized in the ensuing statistical theory (which focused instead on the causally inert properties of probability measures). Nonetheless, both statistical foundations and basic statistics can and should be taught using formal causal models. The causal view of statistical science fits within a broader information-processing framework which illuminates and unifies frequentist, Bayesian, and related probability-based foundations of statistics. Causality theory can thus be seen as a key component connecting computation to contextual information, not "extra-statistical" but instead essential for sound statistical training and applications.

**Acknowledgements**: I am grateful to Steve Cole, Joseph Halpern, Jay Kaufman, Blakeley McShane, and Sherrilyn Roush for their helpful comments on the drafts.

*The only immediate utility of all the sciences is to teach us how to control and regulate future events through their causes*. – Hume [1748]

## Introduction: Scientific Inference *is* a Branch of Causality Theory

I will argue that realistic and thus scientifically relevant statistical theory is best viewed as a subdomain of causality theory, not a separate entity or an extension of probability. In particular, the application of statistics (and indeed most technology) must deal with causation if it is to represent adequately the underlying reality of how we came to observe what was seen – that is, the causal network leading to the data.[1] The network we deploy for analysis incorporates whatever time-order and independence assumptions we use for interpreting observed associations, whether those assumptions are derived from background (contextual) or design information [Pearl 1995, 2009; Robins 2001]. In making this case, I will invoke Pearl's own arguments (e.g., as in Pearl [2009], Wasserstein [2018]) to deduce that statistics should integrate causal networks into its basic teachings and indeed into its entire theory, starting with the probability and bias models that are used to build up statistical methods and interpret their outputs.

Every real data analysis has a causal component comprising the causal network assumed to have created the data set. Decision analysis has a further causal component showing the effects of decisions. Although these causal components are usually left implicit, a primary purpose of design strategies is to rule out alternative causal explanations for observations. Consider one of the most advanced research projects of all time, the search for the Higgs boson. Almost all statistical attention focused on the one-sided 5-sigma detection criterion [Lamb 2012], roughly equivalent to an α-level of 0.0000003, or requiring at least $-\log_2(0.0000003) = 22$ bits of information against the null [Greenland 2019] to declare detection. Yet the causal component is just as important: It includes every attempt to eliminate explanations for such extreme deviations other than the Higgs boson, e.g., the painstaking checks of equipment are actions taken to block the mechanisms that could cause anything near that deviation (other than the Higgs mechanism itself).

[1]This view arguably applies even when dealing with quantum phenomena, at least in the Qbist view [Mermin 2016]. In that view, the laws of quantum mechanics describe how equipment settings causally affect individual perceptions, where the latter become formalized as coherent predictive bets or frequency claims about subsequent observations under those settings (in contrast to other theories that treat quantum probabilities as properties of the environment). Such a controversial view is however unnecessary for the everyday applications of probability and causation that typify most of science and technology, so will not be pursued here.

Thus, because statistical analyses need a causal skeleton to connect to the world, causality is not extra-statistical but instead is a logical antecedent of real-world inferences. Claims of random or "ignorable" or "unbiased" sampling or allocation are justified by causal actions to block ("control") unwanted causal effects on the sample patterns. Without such actions of causal blocking, independence can only be treated as a subjective exchangeability assumption whose justification requires detailed contextual information about absence of factors capable of causally influencing both selection (including selection for treatment) and outcomes [Greenland 1990]. Otherwise it is essential to consider pathways for the causation of biases (nonrandom, systematic errors) and their interactions [Pearl 1995; Greenland et al. 1999; Maclure and Schneeweiss 2001; Hernán et al. 2004; Greenland 2010a, 2012a, 2021].

The remainder of the present paper elaborates on the following points: Probability is inadequate as a foundation for applied statistics, because competent statistical practice integrates logic, context, and probability into scientific inference and decision, using causal narratives to explain diverse data [Greenland et al. 2004]. Thus, given the absence of elaborated causality discussions in statistics textbooks and coursework, we should not be surprised at the widespread misuse and misinterpretation of statistical methods and results. This is why incorporation of causality into introductory statistics is needed as urgently as other far more modest yet equally resisted reforms involving shifts in labels and interpretations for P-values and interval estimates.[2]

As a preliminary, consider that the Merriam-Webster Online Dictionary [2019] defines statistics as "a branch of mathematics dealing with data collection, organization, analysis, interpretation and presentation." Many working statisticians (including me) regard the "branch of mathematics" portion as abjectly wrong, akin to calling physics, computer science or any other heavily mathematical field a branch of mathematics. But we can fix that by replacing "branch of mathematics" with "science" to obtain

> Statistics is the science of data collection, organization, analysis, interpretation and presentation, often in the service of decision analysis.

---

[2]Such as replacement of misleading terms like "statistical significance" and "confidence" by more modest terms like "compatibility" [Amrhein et al. 2019; Greenland 2017b, 2019; Rafi and Greenland 2020; Greenland and Rafi 2020; McShane et al. 2019; Wasserstein et al. 2019].

The amended definition makes no explicit mention of *either* probability or causation, but it is implicitly causal throughout, describing a sequence of actions with at least partial time ordering, each of which is capable of affecting subsequent actions: Study design affects actions during data collection (e.g., restrictions on selection); these actions along with events during data collection (e.g., censoring) affect the data that result; these actions and events affect (or should affect) the study description and the data analysis; and the analysis results will affect the presentation. Overall, the presumed causal structure of this sequence supplies the basis for a justifiable interpretation of the study. Thus, whether answering the most esoteric scientific questions or the most mundane administrative ones, and whether the question is descriptive, causal, or purely predictive, causal reasoning will be crucially involved (albeit often hidden to ill effect in equations and assumptions used to get the "results").

**Causality is central even for purely descriptive goals**

As Pearl has often noted, causal descriptions encode the information and goals that lead to concerns about associations [Pearl 2009]. Consider survey statistics, in which the target question is not itself causal, merely descriptive, such as the proportion of voters who would vote for a given candidate. A competent survey researcher will be concerned about what characteristics C will affect both survey participation (S=1) and voting intent V. Using square brackets to indicate that the observations are conditioned on S=1, this concern is encoded in the diagram

$$[S=1] \leftarrow C \rightarrow V,$$

in which we can see bias in the sampling estimator for the preference distribution Pr(V=v) will be induced by the selection on S. If instead we said only that the concern is about characteristics that are *associated* with both participation and preference (as in S↔C↔V) we would obscure the contextual basis for the concern.

To paraphrase Pearl, statistical analysis without causality is like medicine without physiology. As an example, if we see a difference in ethnic distributions (C) between our survey and population demographic data, we should be concerned about mis-estimating (say) the proportion of Trump voters in the target population. This concern is not because "white ethnicity is *associated* with voting for Trump" as some academic descriptions would have it, but because we expect that being a white male causes sympathy (or prevents antipathy) for Trump's pronouncements relative to being black. That expectation arises from a simple causal relation encoded in C→S, which creates the concern about only seeing preferences of those in the survey, i.e., seeing only Pr(V=v|S=1).

When survey methods attempt to adjust for the difference by reweighting the sample using the target-population ethnicity distribution, that adjustment can be seen as an attempt to counterbalance the C→S arrow in the mechanism generating the sample. This added computation in producing a reweighted sample is traditionally treated as a purely numeric artifice, but is also a causal process: Someone must physically obtain target-weight data and program the reweighting to create the adjusted estimate. It is misleading to describe this action as "simulating removal of an arrow"; it is instead the addition to the data generator of a weighting intervention W in a new causal pathway within

$$[S=1] \leftarrow C \rightarrow V \leftarrow W \leftarrow C$$

W is engineered to (hopefully) balance out the bias from conditioning on selection [S=1]. Note that C appears twice in this diagram to allow it to be written in one line; writing it twice separates the initial effect of C on voter preferences (V) and sample formation (participation S) from its later effect on the analysis weighting W.

**The strength of probabilistic independence demands physical independence**

By data generator, I do *not* mean some abstract structural equation, but rather the entire set of actual physical mechanisms that produce our observations. Even in the simplest games of chance, it is the *physical* (mechanical, causal) independence of coin tosses which licenses our teaching that betting systems for toss sequences will fail to beat simple expectations based on the frequency of heads observed so far. A causal diagram for a sequence of independent identically distributed (i.i.d.) tosses with outcome indicators $Y_1,\ldots,Y_N$ would thus show these N indicators as N isolated (unconnected) nodes.[3] More generally, every missing arrow implies an independence assumption, and such an assumption is really a large *set* of assumptions on the joint distribution of the data $Y_1,\ldots,Y_N$.

One way to measure the information in or logical strength of an independence assumption is by the number of logically independent constraints it imposes (equivalent to the number of parameters whose value it specifies, or the number of dimensions or degrees of freedom it removes from further consideration). Allowing for any possible dependency pattern (as suggested by "nonparametric") among the $Y_1,\ldots,Y_N$ yields a measure of order N factorial; even if we count only pairwise dependencies, the number of patterns is of order $N^2$ (see Appendix 1). Either way, when described honestly, an i.i.d. assumption is not one assumption but rather a *set* of assumptions that grows far faster than the number of observations N. The amount of deductive (digital, syntactical) information in this assumption set is thus beyond anything data frequencies alone could contain; only contextual (background and design) information can supply enough information to warrant such a large set of assumptions.

This enormous logical content of random sampling and randomization illustrates why they are such powerful investigative tools: Only the physical act of blocking all causal effects on selection or treatment can provide deductive justification for the entire set of assumptions corresponding to "independence."

## The Superconducting Supercollider of Selection

In human field studies, realistic causal diagrams should always have a selection (sampling) indicator node S as shown as part of the data-generating process. This node may be influenced by (and perhaps even influence) study variables. By definition, only those with S=1 are observed; thus S will always be conditioned on. If S is affected by more than one variable it will be a conditioned collider and thus a potential bias source under ordinary graphical rules [Greenland 2010a, 2012a]. Most basic causal-diagram introductions (including those I helped write) can be faulted for not emphasizing this fact. We can now fault statistics education for the same reason, in that the "ignorability" of selection under random sampling has led to forgettability of the physical selection mechanism in settings where it is not random in any mechanical sense and thus not ignorable in any practical sense.

[3] A Bayes network would generalize this diagram to show an exchangeable sequence with a node representing the single-toss probability feeding into the $Y_n$.

An important point for graphically representing these problems is that not all of what is known as selection bias arises from S being a collider.[4] For example, classical selection bias requires no collider in the causal graph of data collection. Consider in the earlier voting-survey graph [S=1]←C→V; the bias here corresponds to classical confounding, as it comes from an open back-door path connecting V to S via a shared cause (the causal fork at C). As with confounding, a solution is to condition (stratify) on C, which allows identification of C-conditional voter intentions.

Unlike in classical confounding, however, conditioning is only a partial solution: In the example, the goal is to recover the marginal (C-unconditional) distribution Pr(V=v) of V in the targeted S-unconditional population. Unfortunately, that V marginal is not identified if the graph is the only information available on the target population. This identification is achieved in classical demographic and epidemiologic standardization[5] by averaging the observed C-conditionals Pr(V=v|C=c,S=1) over the C distribution of the target population, Pr(C=c); this procedure assumes however that V is independent of selection given C, so that Pr(V=v|C=c,S=1) = Pr(V=v|C=c), as implied by S←C→V.

A parallel example of selection bias without collider bias arises in studying the effect of a treatment X on an outcome Y when C is a modifier of the treatment effect, as in

[S=1]←C→Y←X

[Hernán 2017]: C is independent of treatment X, and Y is independent of selection S given C, but the S←C→Y path still can bias the estimated marginal X→Y effect given the conditioning on selection (S=1); this bias would become intractable if selection (observation) affected the targeted effects (as in S→Y←X).

## Data and algorithms are causes of reported results

---

[4]This point is contrary to Hernán et al. [2004]; see Hernán [2017] for a reconciliation.

[5]Not to be confused with "standardization" as in dividing a variable by its standard deviation, which damages comparisons of estimates both within and across studies [Greenland et al. 1986, 1991].

The causal sequence continues once the data are collected: A statistical procedure is a data-processing algorithm whose flow chart can be viewed as a causal diagram showing how each computational step determines the next. Usually, each node is a deterministic function of its parents, but may include simulations (as in bootstrap and Markov-Chain Monte-Carlo procedures) that may result in stochastic conditional branches. Finally, the outputs of the algorithm cause researchers and readers to interpret and report the study in particular ways, whether mechanically (e.g., in misreports of “no association” because a P-value exceeded 0.05) or informally, and can strongly affect whether and where the results are published.

Given the causal nature of data generation, calling causal models "extra-statistical" is a misleading characterization of both causality and statistics: Valid statistical analysis is causal to the core; hence, **realistic statistical analysis is a subset of causal analysis**. Not even "extra-distributional" is correct, because the core problem is about factors producing (causing) differences in distributions of those targeted (e.g., voters; patients with a given indication for treatment) and those observed (e.g., survey responders; patients in a trial). Without a causal model for deducing the assumed data distribution from the entire physical data generator, we have no basis for claiming our probability calculations are connected to our target or the world beyond our immediate data.

To summarize so far: Taking off from the Epilogue of Pearl [2009], statistics as conceived and practiced competently is about laying out the causal sequences leading from data to inferences (perceptions) and decisions. Within this sequence, a statistical analysis algorithm or protocol is a causal submodel for how that data will be processed into outputs. Those outputs will then be interpreted as statements connecting the target population to our data under our causally-derived sampling model, with the connections established via open paths in the causal diagram between the target and the data – including connections passing through the ever-present selection node S. Probability plays a central role in terms of formalizing the expected behaviors (propensities) of the data generator under different hypotheses; but that formalization is physically justified only when it is deduced from the causal structure of the generator.

**Getting causality into statistics by putting statistics into causal terms from the start**

Labeling causation as "extra-statistical" creates an excuse to continue to ignore causality theory in statistical teaching and methods research, and stay within the insufficient descriptions of acausal probability theory as the only formal foundation of statistics. That leads to bad practice, such as confusing probabilities of group events with probabilities of individual events within a group. Examples of such confusion [Greenland and Robins 1988; Robins and Greenland 1989; Greenland et al. 2019] may help statisticians recognize causality as an essential component that distinguishes application-relevant statistical theory from acausal probability and its extensions in mathematical statistics. Again, sound applications also need detailed causal explanations of how the data were generated – including the physical mechanisms that led to being in different comparison groups and to inclusion in the data set (S=1).

These causal explanations provide the contextual justifications for the probability models used in the analysis, displaying information about study features that physically constrain data generation. One teaching implication is that students must master causal thinking before they can master real-world statistical inference; thus, basic logic and its causal extensions should be covered from the start of introductory statistics, *before* probability and statistics. But the curriculum for doing so is in its infancy. I used this sequencing in my UCLA courses; however, all incoming students had at least basic statistics, and most also had research methods courses in which at least informal ideas of causality were covered. Thus, the students needed retraining to remove common misconceptions about the implications (or lack thereof) of various statistical results for causal questions.

Students had no trouble mastering the idea of associations passing through causal forks (such as X←C→Y) or mediators (such as X→M→Y); in fact their entire intuition for bias and adjustment came from these two cases. On the other hand, their intuitions for paths though colliders (such as X→S←Y) were backwards, as should be no surprise: Collider bias is by definition the negative or inverse of confounding, because collider bias arises from conditioning (on colliders), whereas confounding is removed by conditioning (on shared causes). Hence, for absolute measures, confounding bias equals the unconditional association minus a conditional association, whereas collider bias equals a conditional association minus the unconditional association.

Again, this view applies not only for causal research questions but also for descriptive survey research. In all real settings in which perfection is unattainable, researchers should try to understand causes of nonresponse, loss, missing data, misreporting, and other sources of uncertainty and inferential distortion[6] - for example by placing these bias sources in a causal diagram to guide study design and interpretation. Only then can they begin to master the far more subtle notions of probabilistic inference from incomplete observations.

**Causation in 20th-century statistics**

Statistical foundation debates raged throughout the last century, but focused exclusively on prioritization of logical criteria such as internal coherence (no violations of the axioms of probability theory) versus self-calibration (meeting select frequency criteria over data sequences generated by the distribution used to derive the data-processing algorithm). Yet formal causal modeling is as old as modern statistical foundations laid down by Fisher, Neyman, DeFinetti and many others in the first half of the 20th century. Although Neyman [1923] went largely unnoticed, potential-outcome ("counterfactual") models entered prestigious statistics journals by the 1930s, and had an ongoing presence before their broad uptake began in the 1980s (e.g. [Welch 1937, Wilk 1955, Copas 1973]). Even without such formalisms, the probability models on which statistical procedures were based were supposed to be frequency summaries of causal mechanisms with certain physical independencies built in by design; these independencies made the mechanisms "ignorable" [Rubin 1978] – a misleading in term insofar as the data-generating mechanism should always be described in detail, never ignored. Such mechanisms include random sampling, which makes selection S an unaffected (exogenous) node, and random allocation, which makes treatment assignment an unaffected node.

Statistical developments in the 20th century were foremost concerned with causal inferences derived from physical randomization, whether by nature as in genetic recombination, or by design. Fisher was often quite straightforward in his causal descriptions and how he regarded causal inference about treatment effects as the central goal of scientific experimentation in the life sciences. By the mid-1930s he had laid out potential outcomes clearly enough (even if only verbally) to see the distinction between the sharp null of no effect on any unit (used to derive randomization tests) and Neyman's weak null of no effect on the mean [Greenland, 1991].

---

[6]These include bad research practices such as "P-hacking": Searching out analyses that give P-values above *or* below a threshold for "significance" [Amrhein et al. 2019; Greenland 2017a, 2017b, 2019].

His *Design of Experiments* [Fisher 1935] gives primacy to experimental action (design) over mathematics, as seen in sec. 2 of his introduction to the first edition, in which he states

> "I have assumed, as the experimenter always does assume, that it is possible to draw valid inferences from the results of experimentation; **that it is possible to argue from consequences to causes**, from observations to hypotheses; as a statistician would say, from a sample to the population from which the sample was drawn, or, as a logician might put it, from the particular to the general."

His ensuing verbal descriptions were soon formalized by others into a clear potential-outcome model form, where *for each unit* explicit counterfactual (unobserved) treatment assignments lead to possibly distinct outcomes (e.g., see Welch [1937, p. 22-23]).

Nonetheless, the statistical theory that dominated subsequent advanced teaching and methods research became an extension of measure-theoretic probability, a development decried by those who followed Fisher in emphasizing the importance of context [Box 1990]. It is thus somewhat ironic that Fisher's downfall (as manifested in his defense of smoking against charges of carcinogenicity) was his inability to neutrally synthesize all available evidence sources, particularly in mishandling sources of information not derived from physical randomization. This failing can be viewed as one of being unable to form realistic models for confounding effects coupled with (or perhaps caused by) by personal wishes for vindication of his own smoking habit [Stolley 1991]. These sorts of "human factors" are themselves extraneous causes of what gets reported and publicized, and thus need to be accounted for in any realistic model for literature analysis [Greenland 2012b, 2017a, 2017b].

## Causal analysis vs. traditional statistical analysis

In applied statistics, assumptions are made to simplify modeling effort, which like everything else is resource constrained. For example, the standard modeling assumption "linear in the natural parameter" is rarely if ever deduced from anything; instead, standard statistical methods treat it as certainly true provided there is no evidence to contrary (even if there is little evidence to judge its accuracy or practical impact). This convention is based on the ease of use of such models, especially their transparency and computational stability relative to intrinsically nonlinear models, along with the idea that basic linear trend components are sometimes the only components that are needed or that can be stably estimated from available data.

A retreat from causal to convenience justification is only to be expected when applications involve complexities beyond complete formal (algorithmic) modeling capacities, as in biology, medicine, and social sciences. In such applications, all models are wrong at some practical level of analysis, and are often wrong in very consequential ways *even when they are useful for improving predictions of yet-unseen events such as treatment effects*. The classic epidemiologic example is malaria, a disease whose name means "bad air" in the parent Italian. Before modern times, social groups noted that malaria rates were higher near swamps and attributed that to toxic effects on the air from the swamps, as suggested by the foul smell associated with swamps. This wrong theory (causal-system model) of

swamp→ toxic air → malaria

housing location → toxic air → malaria

led to successful interventions such as draining swamps and building elevated houses, even though it missed the actual causal structure of

swamp → mosquito exposure → malaria

housing location → mosquito exposure → malaria which predicted the same intervention effects. To explain these successes of the wrong model, we may note that the swamp intervention tested only the swamp→malaria effect while the housing intervention tested only the housing→malaria effect. Both interventions left wide open the identity of the intermediates (and thus specifics of the mechanism for intervention), yet were taken to demonstrate the (in-fact untested) pathway of toxic air.

Such examples show that causal theories can include important mistakes even while successfully predicting intervention effects, and show why those theories should not be taken as true because of such successes (even in a world where causal laws are stable and thus inductive reasoning is justified). They instead need ongoing novel tests (not just "replication") before basing actions on pathways that have not yet been tested by experiments. The enhanced risk of error for a mechanistic causal theory over a mere predictive/associative theory is not a disadvantage, however: it reflects the greater specificity, greater logical content, and hence greater testability of such theories, properties which are often promoted as hallmarks of good scientific theories (Popper 1962).

That such a theory can pass apparently strong experimental tests yet be erroneous in important ways (as in the malaria example) is one reason pragmatic analysts reject notions of "experimental support" for scientific (real-world) causal theories. Other theories (including many never imagined) may pass the same experimental test, so at most we can only say an experiment supports the broad class of theories which predict results close to what was observed. Put another way: An intervention experiment provides evidence only on *classes* of mechanisms (those whose diagrams have directed paths from the observed intervention to the observed outcome), not specific mechanisms, and thus leaves open many details of intervention effects.

That caution applies even more strongly in passive observational (nonexperimental) studies, especially when their data are "analyzed" (summarized) by statistics based on randomization assumptions. In that case one can view a conventional interval estimate as a blur around the point estimate indicating irreducible uncertainty about the behavior of the data generator. But any inferential connection of these summaries to a targeted treatment effect should be mediated by explicit causal models; specifically, extraction of information about the target effect (e.g., in form of credible uncertainty intervals for the target) requires causal models for physical data generation that include nonrandom variation (bias) sources beyond the treatment [Greenland 1990, 2012a; Greenland et al. 1999; Maclure and Schneeweiss, 2001; Robins 2001; Hernán et al. 2004; Glymour and Greenland 2008]. It also requires recognition that effects cannot always be identified by observed associations, and that some effects cannot be statistically identified at all, even from randomized trials [Kaufman 2009; Robins and Richardson 2011].

**Relating causality to traditional statistical philosophies and "objective" statistics**

As has been long and widely emphasized in various terms (e.g., [Cox 1978; Box 1980, 1990; Rubin 1984; Good 1992; Barnard 1996; Chatfield 2002; Kelly and Glymour 2004; Greenland 2006, 2010b; Senn 2011; Gelman and Shalizi 2013], frequentism and Bayesianism are incomplete both as learning theories and as philosophies of statistics, in the pragmatic sense that each alone are insufficient for all sound applications. Notably, causal justifications are the foundation for classical frequentism, which demands that all model constraints be deduced from real mechanical constraints on the physical data-generating process. Nonetheless, it seems modeling analyses in health, medical, and social sciences rarely have such physical justification.

Beyond graphs, causality theory formalizes design information (such as randomization and matching) by the constraints that information places on the distributions of unobserved variables (e.g., [Greenland 1990; Pearl 1995; Robins 2001; Hernán and Robins 2020]). Use of that information is especially important when the modeled data generator is not fully understood as a coherent whole – a problem long recognized and discussed at length in the literature on model uncertainty (e.g., [Leamer 1978; Box 1980]). The deficiency of strict coherent (operational subjective) Bayesianism is its assumption that all aspects of this uncertainty have been captured by the prior and likelihood, thus excluding the possibility of model misspecification [Leamer 1978; Box 1980; Senn 2011]. DeFinetti himself was aware of this limitation:

> "…everything is based on distinctions which are themselves uncertain and vague, and which we conventionally translate into terms of certainty only because of the logical formulation…In the mathematical formulation of any problem it is necessary to base oneself on some appropriate idealizations and simplification. This is, however, a disadvantage; it is a distorting factor which one should always try to keep in check, and to approach circumspectly. It is unfortunate that the reverse often happens. One loses sight of the original nature of the problem, falls in love with the idealization, and then blames reality for not conforming to it." [DeFinetti 1975, p. 279][7]

By asking for physically causal justifications of the data distributions employed in statistical analyses (whether those analyses are labeled frequentist or Bayesian), we may minimize the excessive certainty imposed by simply assuming a probability model and proceeding as if that idealization were a known fact.

DeFinetti was of course writing in support of a contentious, purely subjective view of probability, and the utility of the entire "subjective"/"objective" distinction in statistics has been questioned [Gelman and Hennig 2017]. Nonetheless, many statisticians assign primacy to "objective" model components (those derivable from observed mechanisms, such as random-number generators). What supports a claim that a variable is "completely random" (fully randomized) in an objective frequency sense? Modern causality theory can identify this randomness with the assumption that the variable is exogenous or instrumental, in that its causes affect the system under study only through the variable [Pearl 2009]. Again, in "objective" theory this sharp, strong assumption is *deduced* from the physical data-generating mechanism, not from observed frequencies or other purely associational information.

[7] I am indebted to Stephen Senn for reminding me of this and other remarkable passages in DeFinetti.

Consider "fair" coin tossing, in which the influence of the person tossing (who might be a magician) is blocked by having them throw the coin against a wall and then step back before the bounce and landing, thus blocking skilled tossing and other trickery as influences of the outcome. Then, even under classical deterministic mechanics, the functional complexity of the relation of the outcome to the initial toss is transcomputable or chaotic. This type of complexity forces our predictions to rely on distributions that arise as attractors of statistical behavior (e.g., laws of large numbers, central-limit effects), instead of deterministic mathematical models. In doing so we assume a certain causal stability across trials whose consequences are summarized in our models. Such a stability assumption needs justification based on direct observation (the physical mechanism is unchanging) and thus is objective; without that, causal stability is an underived (and usually implicit) assumption and thus is not objective in this sense.

In this way, the traditional “objective”/”subjective” distinction in statistical methods resides within causality theory, not in the "frequentist" vs "Bayesian" distinction (which are both vague labels for highly heterogeneous collections of statistical tools and philosophies, as Good [1971] explained for Bayes). The core idea behind “objective” statistics is that one demands that each distribution used in the statistical processing of the data be derivable from a verifiable physical (causal) mechanism. That demand can be made regardless of whether that processing is labeled “frequentist”, “Bayesian”, “likelihoodist” or something else, a view which does not exclude Bayesian methods, but does reject mere expressions of opinions as priors for those methods [von Mises 1981].

## Discussion

Judea Pearl has been a celebrated promoter of causal models over pure probability, especially for encoding the background (contextual) information in a problem [Pearl 1995, 2001, 2009]. At times however he has referred to causality as “extra-statistical,” a label which ignores the realities that any applied statistician must face in practice. Those realities make causality integral to statistics; yet, by calling causality “extra-statistical” we absolve those bearing the professional label “statistician” of any responsibility to understand let alone teach causality theory. Fortunately, many younger statisticians have a keen interest in causal models as tools to create better statistical science. To encourage this trend we should include causal models from the start of statistical training as an integral component of study design and data analysis – in addition to complementary presentation of frequentist and Bayesian ideas.

As a less-often stated yet even more fundamental need, basic statistics should begin with the elements of deductive logic. When I was teaching statistical foundations and principles, most students I encountered (including statistics majors) had neither studied nor fully understood basic logical principles, and thus were prone to naïve fallacies in verbal arguments. Thus the topic sequence in my class covered logic as a foundation for causal thinking, followed by causality theory as a foundation for probability and association explanation. This material was contrasted to their previous instruction, which typically involved rote application of mysterious descriptions and formulas for statistical comparisons and regressions. Students were always delighted to at last see applied statistics as the coordinated merging of the three essentials of logic, causation, and probability to provide a transparent foundation for sound study design, analysis, and interpretation.

Admittedly, traditionally trained statisticians may be too firmly wedded to probabilistic foundations to ever concede this causal primacy, and some radical subjective Bayesians reject causality altogether (e.g., Lad [2006]). Nonetheless, probabilists curious about the causal approach may more easily conceive the unification of causality and probability within information theory, which can serve as an overarching framework for statistical modeling and inference (I have found that an information framework even helps students correctly understand P-values [Greenland 2019; Rafi and Greenland 2020]). Causal diagrams then provide an intuitive representation of information flows as time-sequential functional relations across event sequences.

**Conclusion**

Statistical science (as opposed to mathematical statistics) involves far more than data – it requires realistic *causal* models for the generation of that data and the deduction of their empirical consequences. Evaluating the realism of those models in turn requires immersion in the subject matter (context) under study. Decisions further require explication of the various pathways by which those decisions would cause gains (benefits) and losses (costs). Bringing these causal elements to the foreground is essential for sound teaching and applications of statistics.

**Appendix. A counting measure for the logical content of a finite exchangeability assumption**

For any formal deductive system and set of constraints A in the system, define A as logically minimal if it satisfies the joint deductive independence condition: For any pair (B,C) of disjoint nonempty subsets of A, C cannot be deduced from B. We may then define the logical-content measure ν(G) of an arbitrary set of constraints G in the system as the smallest cardinality |A| among minimal subsets A of G; ν(G) may be infinite if G is infinite.

Now consider the common statistical assumption that the observations $Y_1,\ldots,Y_N$ are independent identically distributed conditional on any model m in a set M. Then, given a prior distribution on M, the $Y_1,\ldots,Y_N$ are unconditionally exchangeable; that is, every one of the N! permutations of indices in the joint distribution leaves that distribution unchanged. Exchangeability is logically equivalent to N!-1 independent constraints, one for each non-null permutation; denoting the set of these constraints by G, with no further constraint we have ν(G) = |G| = N!-1. By imposing further constraints on the joint distribution we may reduce ν(G) considerably. Nonetheless, even with the extreme simplification of multivariate normality we get ν(G) of order $N^2$ (since exchangeability requires homogeneous variances and homogeneous covariances), and thus still entails far more constraints than there are observations N.